\documentclass[pra,
                preprint,
                amsmath,
                amssymb,
                showpacs,
                superscriptaddress]{revtex4}
\usepackage{graphicx} 
\usepackage{dcolumn}  
\usepackage{bm}       
\usepackage{amsfonts}
\usepackage{dsfont}
\usepackage{color}

\begin{document}

\title{Time-resolved ultrafast x-ray scattering from an incoherent electronic mixture}

\author{Gopal Dixit}
\email[]{gdixit@phy.iitb.ac.in}
\affiliation{%
Department of Physics, Indian Institute of Technology Bombay,
            Powai, Mumbai 400076  India}

\author{Robin Santra}
\email[]{robin.santra@cfel.de}
\affiliation{%
Center for Free-Electron Laser Science, DESY,
            Notkestrasse 85, 22607 Hamburg, Germany }
\affiliation{%
The Hamburg Centre for Ultrafast Imaging,
Luruper Chaussee 149, 22761 Hamburg, Germany }
\affiliation{%
Department of Physics, University of Hamburg, 20355 Hamburg,
Germany}

\date{\today}



\begin{abstract}
Time-resolved ultrafast x-ray scattering from photo-excited 
matter is an emerging method to image ultrafast dynamics 
in matter with atomic-scale spatial and temporal resolutions.
For a correct and rigorous understanding of current and upcoming imaging experiments, 
we present the theory of time-resolved x-ray scattering from an incoherent electronic mixture
using quantum electrodynamical theory of light-matter interaction.  
We show that the total scattering signal is an incoherent sum of the individual scattering 
signals arising from different electronic states and therefore  
heterodyning of the individual  signals is not possible for an ensemble of gas-phase 
photo-excited molecules. 
We scrutinize the information encoded in the total signal 
for the experimentally important situation 
when pulse duration
and coherence time  of the x-ray pulse are short in comparison to the 
timescale of the vibrational motion
and long in comparison to the
timescale of the electronic motion, respectively. 
Finally, we show that 
in the case of an electronically excited crystal
the total scattering signal imprints the interference 
 of the individual  scattering amplitudes  associated with different electronic states 
 and heterodyning is possible.  
 \end{abstract}

\maketitle
\section{Introduction}
These days, technological advancements make it possible to generate 
tunable,  ultrafast and intense x-ray pulses from x-ray free-electron lasers 
(XFELs)~\cite{emma2, ishikawa2012, pellegrini2016physics}. 
Several interesting successful experiments have been  performed for 
systems  ranging from atoms  to complex biomolecules  since the beginning of 
the operation of the first XFEL 
in the hard-x-ray regime, the Linac Coherent Light Source~\cite{bostedt2016linac}. 
X-ray scattering is an indispensable method to obtain 
real-space structure of matter with atomic-scale spatial resolution~\cite{Nielsen}. 
For a complete 
understanding of the structural-functional relationship of matter, it is crucial 
to image the dynamical evolution of matter in action with their natural  spatial and 
temporal resolutions~\cite{gaffney2007imaging}.
These ultrashort x-ray pulses from XFELs (LCLS, SACLA, European XFEL) 
offer femtosecond temporal resolution for structural imaging of matter. 
The pump-probe approach is a common way to trace ultrafast dynamics, 
where the dynamics are initiated by a short pump pulse and the activated dynamics 
are subsequently interrogated by the probe pulse at a precise time.    
Time-resolved x-ray scattering (TRXS) from 
temporally evolving matter  
records molecular ``movies'' to map out the atomic and electronic motions
on their natural timescale~\cite{bucksbaum2007, peplow2017next}. 

Imaging of aligned gas-phase molecules  
has been demonstrated experimentally at LCLS~\cite{kupper2014x}.   
Subsequently,  a ring opening electrocyclic chemical reaction in cyclohexadiene 
was probed using TRXS. 
The temporally evolving structural information was
extracted by a comparison of experimental data with theoretical simulations  
~\cite{minitti2015imaging}.  
Also, cis/trans structural dynamics in photoactive yellow protein was imaged via TRXS~\cite{pande2016femtosecond}.   
Recently, Glownia and co-workers  imaged 
ultrafast vibrational motion in  photo-excited molecular iodine at LCLS using TRXS 
with a spatial and temporal  resolution of 0.3 \AA ~ and 30 femtoseconds, respectively. 
Moreover, the idea of holographic (heterodyne) detection, 
based on the assumption of interference between the ground-state stationary charge distribution and the nonstationary excitation, was used to analyze the data in this experiment
~\cite{glownia2016self}. 
Also, plasma  based ultrashort x-ray sources   
have been used to image various non-equilibrium phenomena in matter using TRXS~\cite{elsaesser2014perspective}. 
These state-of-the-art experiments  have demonstrated that TRXS is   
an emerging and promising approach to image ultrafast processes in 
real-space and in real-time with atomic-scale spatio-temporal resolutions. 

When a pump pulse interacts with a molecule or a crystal, it may 
create a superposition of electronic states.  
Depending upon the parameters of the pump, the superposition is a (partially) 
coherent or incoherent
mixture of electronic states. 
If the  pump pulse fluctuates from shot to shot
(for instance, there is typically no carrier envelope phase stabilisation), 
an incoherent electronic superposition is created.
An incoherent electronic ensemble  in molecular iodine was created
in the recent experiment by  Glownia and co-workers~\cite{glownia2016self}.  
There was no coherence between the ground and excited electronic states 
and only full or partial vibrational coherence in each electronic state was present.
The notion of heterodyne detection used to 
interpret the TRXS  signal~\cite{glownia2016self} was recently challenged by Mukamel and co-workers  in a brief comment ~\cite{bennett2017comment}.  
Note that the concept of heterodyne detection has been used to interpret the TRXS 
signal from an electronically excited solid~\cite{elsaesser2014perspective, zamponi2012ultrafast}, but there is no 
consistent quantum theory of TRXS from a nonstationary  solid showing  that 
the heterodyne detection is feasible. 
So why does the concept of heterodyne detection
come under debate  in the case of  photo-excited gas-phase molecules 
while it is assumed to work well in the case of 
photo-excited  crystal as has been used by Elsaesser and co-workers for a long time~\cite{elsaesser2014perspective, zamponi2012ultrafast}? 
In this article, we will explore this question and will show that 
even though in both cases the scattering signal is based on 
the same mathematical structure,  the 
information encoded in the respective signals is completely different. 
Note that the possibility of heterodyne detection in TRXS from a coherent electronic wave packet
in an atom  has been proposed by Vrakking and Elsaesser~\cite{vrakking2012}. 
A time-resolved version of phase contrast imaging has been proposed to image 
a coherent electronic wave packet~\cite{dixit2013prl}.

The purpose of the present paper is to provide a rigorous theoretical  analysis
of TRXS from an incoherent electronic mixture, explicitly taking into account 
molecular vibrations (lattice dynamics in the case of 
a solid)  using the quantum theory of light-matter 
interaction, and to discuss 
under what conditions heterodyne detection is possible.   
To the best of our knowledge, this is the first quantum-electrodynamic-based derivation or maybe even the first derivation within any framework for TRXS from an incoherent electronic mixture.
Note that the theory of TRXS from a coherent electronic wave packet 
is well-established at different levels  
and the information encoded in the scattering signal has been discussed in detail ~\cite{cao1998ultrafast, bratos2002, bratos2004, henriksen, lorenz2010theory, dixit2012, dixit2013jcp, dixit2014theory, bredtmann2014x, bennett2014time, santra2014}. 
This paper is structured as follows. 
Section II
elucidates the theoretical framework required to describe TRXS from an incoherent electronic wave packet.
Section III presents results and a discussion of the theory presented in the previous section. 
Section III is sub-divided into two subsections, where we present: 
A) an analysis of the theory in the situation when the x-ray pulse duration is much shorter than the dynamical timescale of the nuclear motion in a molecule, and 
the interplay of coherence time and pulse duration of the x-ray pulse in comparison to the characteristic dynamical timescale in a molecule;   
and 
B) TRXS from  an electronically excited  crystal. Conclusions are presented  in Sec. IV. 
   
\section{Theory} \label{secII}
In this article, atomic units are used throughout  unless specified otherwise. 
The density matrix for an incoherent ensemble prepared by the pump pulse reads
\begin{equation}\label{eq01}
\hat{\rho}^{\textrm{m}}_{in}(t_{0}) = \sum_{I} {p_{I}}  |\phi_{I}(\mathbf{R}) \rangle  | \chi_{I}(t_{0}) \rangle \langle \phi_{I}(\mathbf{R})| \langle \chi_{I}(t_{0})|. 
\end{equation} 
Here,
 $ \hat{H}_{\textrm{el}}(\mathbf{R})  | \phi_{I}(\mathbf{R})  \rangle  = E_{I}(\mathbf{R}) | \phi_{I}(\mathbf{R})  \rangle $ where
 $ \hat{H}_{\textrm{el}}(\mathbf{R})$ is the electronic Hamiltonian 
and $E_{I}(\mathbf{R})$ is the potential energy surface of the $I-$th electronic state
within the Born-Oppenheimer approximation. $p_{I}$ is the probability to find the molecule 
in the electronic state $| \phi_{I}(\mathbf{R}) \rangle$.  
Both $| \phi_{I}(\mathbf{R}) \rangle$ and $E_{I}(\mathbf{R})$ depend 
on the nuclear coordinates $\mathbf{R}$ parametrically.  In Eq.~(\ref{eq01}), 
the vibrational wave packet  in the $I$-th electronic state is given by
\begin{equation}\label{eq14}
| \chi_{I} (t_{0}) \rangle =  \sum_{\xi} C_{I; \xi} e^{-i E_{I; \xi}t_{0}} | \chi_{I; \xi} \rangle, 
\end{equation} 
where $ | \chi_{I; \xi} \rangle$ obeys the nuclear Schr{\"o}dinger equation, 
 $[\hat{T}_{\textrm{N}} + E_{I}(\mathbf{R})] | \chi_{I;\xi}  \rangle = E_{I;\xi} | \chi_{I;\xi} \rangle$, 
 with $\hat{T}_{\textrm{N}}$  the nuclear kinetic energy operator. 
The form of the density matrix presented in Eq.~(\ref{eq01}) implies that  
there is no coherence between any electronic states, but 
perfect vibrational coherence in each electronic state. 

Within the language of quantum field theory, a consistent quantum theory 
for the matter and radiation fields is applied.
The light-matter interaction Hamiltonian, from the principle of  
minimal-coupling in the Coulomb gauge, is  
~\cite{craig1984} 
\begin{equation}\label{eq1}
\hat{H}_{\textrm{int}} = \alpha \int d^{3}x
~\hat{\psi}^{\dagger}(\mathbf{x})~
\left[\hat{\mathbf{A}}(\mathbf{x}) \cdot
\frac{\boldsymbol{\nabla}}{i}\right] ~\hat{\psi}(\mathbf{x}) +
\frac{\alpha^{2}}{2}\int d^{3}x ~\hat{\psi}^{\dagger}(\mathbf{x})~
\hat{\mathbf{A}}^{2} (\mathbf{x})~\hat{\psi}(\mathbf{x}),
\end{equation}
where 
$\hat{\psi}(\mathbf{x})~[\hat{\psi}^{\dagger}(\mathbf{x})]$ is the
annihilation [creation] field operator for an electron at position
$\mathbf{x}$, $\alpha$ is the fine-structure constant,
$\hat{\mathbf{A}}$ is the vector potential operator of the
radiation and $\frac{\boldsymbol{\nabla}}{i}$ is the canonical
momentum of an electron. 
In the present formalism,  
we only focus on scattering induced by the  $\hat{\mathbf{A}}^{2} $ operator  
and will not consider the contribution in the scattering process
from the $\hat{\mathbf{A}}(\mathbf{x}) \cdot \boldsymbol{\nabla}$ 
term given in Eq.~(\ref{eq1}), i.e., we neglect the contribution
from the dispersion correction.   
In general, the radiation field must be considered 
as a statistical mixture of photons occupying all possible
electromagnetic modes and $\hat{\mathbf{A}}$ is written as 
~\cite{craig1984}
\begin{equation}\label{eq2}
\hat{\mathbf{A}}(\mathbf{x}) = \sum_{\mathbf{k}, s} \sqrt{\frac{2
\pi}{V \omega_{\mathbf{k}} \alpha^{2}}} \left\{
{\hat{a}_{\mathbf{k}, s}} \boldsymbol{\epsilon}_{\mathbf{k}, s}
e^{i \mathbf{k} \cdot \mathbf{x}} + \hat{a}^{\dagger}_{\mathbf{k},
s} \boldsymbol{\epsilon}^{*}_{\mathbf{k}, s} e^{-i \mathbf{k}
\cdot \mathbf{x}} \right \}.
\end{equation}
Here, $\omega_{\mathbf{k}}$ is the
energy of a photon in the $\mathbf{k}$-th mode  
and $V$ is the quantization volume. 
$\boldsymbol{\epsilon}_{\mathbf{k}, s}$ is the
polarization vector in the $\mathbf{k}, s$ mode and 
$\hat{a}^{\dagger}_{\mathbf{k},s} ~
(\hat{a}_{\mathbf{k},s})$ is the photon creation (annihilation)
operator with $\mathbf{k}$ being
 the wave vector and $s$ the polarization index of a given
mode.
The initial density operator of the radiation field is written as
~\cite{loudon1983}
\begin{equation}\label{eq3}
\hat{\rho}^{X}_{in}=\sum_{\{n\},\{\bar{n}\}}
\rho^{X}_{\{n\},\{\bar{n}\}} |\{n\}\rangle \langle\{\bar{n}\}|,
\end{equation}
with $\rho^{X}_{\{n\},\{\bar{n}\}}$ denoting the populations and
coherences of all the occupied field modes associated with the
incoming beam. ${\{n\}}$ denotes a complete set of numbers that
specify the number of photons in all field modes. All the
scattering modes are unoccupied in $\hat{\rho}^{X}_{in}$.

The differential scattering probability (DSP), a crucial quantity in
x-ray scattering, is expressed as 
\begin{equation}\label{eq4}
\frac{dP}{d\Omega} =
\frac{V\alpha^{3}}{(2\pi)^{3}}\int_{0}^{\infty}
d\omega_{\mathbf{k}_{s}} \omega_{\mathbf{k}_{s}}^{2} ~
P(\mathbf{k_{s}}),
\end{equation}
where $\omega_{\mathbf{k}_{s}}$ refers to the scattered photon energy and
$P(\mathbf{k_{s}})$  is the probability of observing a scattered photon with momentum 
$\mathbf{k_{s}}$, which is different from the momenta of the incoming photons. 
Using the expression for $P(\mathbf{k_{s}})$ given in our earlier work~\cite{dixit2012}, 
the DSP can be written   as
\begin{equation}\label{eq5}
\frac{dP}{d\Omega} =
\frac{V\alpha^{3}}{(2\pi)^{3}}\int_{0}^{\infty}
d\omega_{\mathbf{k}_{s}} ~ \omega_{\mathbf{k}_{s}}^{2}~W_{\Delta
\omega}({\omega_{\mathbf{k}_{s}}}) ~ \sum_{J; \mu} 
\sum_{\{n^{\prime}\}} \langle \chi_{J; \mu} | \langle \phi_{J}( \mathbf{R});
\{n^{\prime}\}|\hat{\rho}_{f}|\phi_{J}(\mathbf{R}); \{n^{\prime}\} \rangle | \chi_{J; \mu}\rangle, 
\end{equation}
where $W_{\Delta \omega}({\omega_{\mathbf{k}_{s}}})$ is 
a spectral window function used to model the
range of scattered photon energies accepted by the detector;
 $\hat{\rho}_{f}$ is the density operator of the entire system at
the time of measurement,
\begin{equation}\label{eq6}
\hat{\rho}_{f} = \lim_{t_{f} \to \infty} \lim_{t_{0} \to -\infty}
\hat{U}_{\textrm{total}}(t_{f}, t_{0}) \hat{\rho}^{s}_{in}
\hat{U}_{\textrm{total}}^{\dagger}(t_{f}, t_{0}),
\end{equation}
where $\hat{\rho}^{s}_{in} = \hat{\rho}^{\textrm{m}}_{in} \otimes
\hat{\rho}^{X}_{in}$, and $\hat{U}_{\textrm{total}}(t_{f},t_{0})$
is the time-evolution operator for the whole system, matter and x rays. 

First-order time-dependent perturbation theory 
is used to compute $\hat{\rho}_{f}$:
\begin{eqnarray}\label{eq7}
\hat{\rho}_{f} & = & \lim_{t_{f} \to \infty} \lim_{t_{0} \to
-\infty} \int_{t_{0}}^{t_{f}} \int_{t_{0}}^{t_{f}} dt_{1} dt_{2} 
\sum_{\{n\},\{\bar{n}\}} \rho^{X}_{\{n\},\{\bar{n}\}}  \nonumber \\ 
&& \times  \sum_{I} {p_{I}}   
\left [ \hat{U}_{\textrm{m, x}}(t_{f}, t_{1}) ~
\hat{H}_{\textrm{int}} ~ \hat{U}_{\textrm{m, x}}(t_{1}, t_{0})  ~
 | \chi_{I}(t_{0}) \rangle |\phi_{I} \rangle 
 |{\{n\}} \rangle \langle {\{\bar{n}\}} |
  \right. 
  \nonumber \\
& & \times \left.  \langle \phi_{I}| 
\langle \chi_{I}(t_{0})| ~
\hat{U}_{\textrm{m, x}}^{\dagger}(t_{2}, t_{0})  ~
\hat{H}_{\textrm{int}}^{\dagger} ~
\hat{U}_{\textrm{m, x}}^{\dagger}(t_{f}, t_{2}) \right].
\end{eqnarray}
Here, $\hat{U}_{\textrm{m, x}}$ is the time-evolution operator of the noninteracting matter and x-ray fields. 
The position of the nuclei, $\mathbf{R}$, has been dropped  to make the expressions more compact.  
 
After substituting the result for $\hat{\rho}_{f}$ from Eq.~(\ref{eq7}) 
 in Eq.~(\ref{eq5}),  we obtain for the DSP 
\begin{eqnarray}\label{eq8}
\frac{dP}{d\Omega} & = & 
\lim_{t_{f} \to \infty}  \sum_{s_{s} = 1}^{2} \sum_{\mathbf{k}_{1}s_{1}\mathbf{k}_{2}s_{2}} \sqrt
{\frac{ \pi^{2}}{V^{2}\omega_{\mathbf{k}_{1}}}}
\sqrt
{\frac{  \pi^{2}}{V^{2}\omega_{\mathbf{k}_{2}}}}
\frac{V\alpha^{3}}{(2\pi)^{3}}
(\boldsymbol{\epsilon}_{\mathbf{k}_{1},s_{1}} \cdot
\boldsymbol{\epsilon}^{*}_{\mathbf{k_{s}},s_{s}})  
(\boldsymbol{\epsilon}^{*}_{\mathbf{k}_{2},s_{2}} \cdot
\boldsymbol{\epsilon}_{\mathbf{k_{s}},s_{s}})  \nonumber \\
&& \times  \int_{0}^{\infty} d\omega_{\mathbf{k}_{s}} \omega_{\mathbf{k}_{s}} ~ W_{\Delta
\omega}({\omega_{\mathbf{k}_{s}}}) ~ 
\int_{-\infty}^{t_{f}} \int_{-\infty}^{t_{f}} dt_{1} dt_{2} ~
 \nonumber \\
& & \times  \sum_{J; \mu} \sum_{I} {p_{I}}   
\Biggl[  \int d^{3}x \langle \chi_{J; \mu} |  \langle \phi_{J} | ~\hat{U}(t_{f}, t_{1}) ~
~ \hat{\psi}^{\dagger}(\mathbf{x})~ \hat{\psi}(\mathbf{x}) ~ |\phi_{I} \rangle 
 | \chi_{I} , t_{1} \rangle  
 e^{i(\mathbf{k}_{1}-\mathbf{k}_{s}) \cdot \mathbf{x}}
 \Biggr.   \nonumber \\
&& \times 
\Biggl.    \int d^{3}x^{\prime}
\langle \chi_{I}, t_{2} |  \langle \phi_{I} | 
~\hat{\psi}^{\dagger}(\mathbf{x}^{\prime})~
\hat{\psi}(\mathbf{x}^{\prime})~
~ \hat{U^{\dagger}}(t_{f}, t_{2})~ | \phi_{J} \rangle  | \chi_{J; \mu} \rangle  
e^{-i(\mathbf{k}_{2}-\mathbf{k}_{s}) \cdot \mathbf{x}^{\prime}} \Biggr] \nonumber \\
&& \times  \Biggl[ \sum_{\{n^{\prime}\}} \sum_{\{n\}, \{\bar{n}\}}
\rho^{X}_{\{n\},\{\bar{n}\}}  \langle \{n^{\prime}\} |
{\hat{a}_{\mathbf{k}_{1}, s_{1}}}
\hat{a}^{\dagger}_{\mathbf{k}_{s}, s_{s}} +
\hat{a}^{\dagger}_{\mathbf{k}_{s}, s_{s}}
{\hat{a}_{\mathbf{k}_{1}, s_{1}}} | \{n\} \rangle
~ e^{i(E_{\{n^{\prime}\}}-E_{\{n\}})t_{1}} \Biggr.   \nonumber \\
&& \Biggl.  \times
e^{i(E_{\{\bar{n}\}}-E_{\{n^{\prime}\}})t_{2}} \langle \{\bar{n}\}
| {\hat{a}_{\mathbf{k}_{s}, s_{s}}}
\hat{a}^{\dagger}_{\mathbf{k}_{2}, s_{2}} +
\hat{a}^{\dagger}_{\mathbf{k}_{2}, s_{2}}
{\hat{a}_{\mathbf{k}_{s}, s_{s}}} | \{n^{\prime}\} \rangle
\Biggr]. 
\end{eqnarray}
Here, $\hat{U}$ is the time-evolution operator
associated with the matter Hamiltonian, and
$E_{\{n\}}$ is the energy corresponding to Fock state
$|{\{n\}}\rangle$. 

Let us rearrange different terms in above equation, so that we can write the 
expression for the DSP  as
\begin{eqnarray}\label{eq9}
\frac{dP}{d\Omega} & = &
\sum_{s_{s} = 1}^{2} ~ \sum_{\mathbf{k}_{1}s_{1}\mathbf{k}_{2}s_{2}} ~
\frac{\alpha^{3}}{2\pi V
\sqrt{\omega_{\mathbf{k}_{1}}\omega_{\mathbf{k}_{2}}}}
 (\boldsymbol{\epsilon}_{\mathbf{k}_{1},s_{1}} \cdot \boldsymbol{\epsilon}^{*}_{\mathbf{k_{s}},s_{s}})
 (\boldsymbol{\epsilon}^{*}_{\mathbf{k}_{2},s_{2}} \cdot \boldsymbol{\epsilon}_{\mathbf{k_{s}},s_{s}})  
\int_{0}^{\infty} d\omega_{\mathbf{k}_{s}} \omega_{\mathbf{k}_{s}} 
~ W_{\Delta \omega}({\omega_{\mathbf{k}_{s}}})
\nonumber \\
& & \times \int_{-\infty}^{\infty} \int_{-\infty}^{\infty} dt_{1} dt_{2}   
 \sum_{I} {p_{I}}  \int d^{3}x ~ \int d^{3}x^{\prime} 
\left\langle \chi_{I}, t_{2} \left| \left\langle \phi_{I} \left|  ~\hat{n}(\mathbf{x}^{\prime}) ~
\hat{U}(t_{2}, t_{1}) ~\hat{n}(\mathbf{x}) ~ \right|\phi_{I} \right \rangle
\right | \chi_{I}, t_{1} \right\rangle    \nonumber \\
&& \times  e^{-i(\mathbf{k}_{2}-\mathbf{k}_{s}) \cdot
\mathbf{x}^{\prime}} ~ e^{i(\mathbf{k}_{1}-\mathbf{k}_{s}) \cdot
\mathbf{x}} ~ \textrm{Tr}[\hat{\rho}^{X}_{in}
{\hat{a}^{\dagger}_{\mathbf{k}_{2}, s_{2}}}
{\hat{a}_{\mathbf{k}_{1}, s_{1}}}] ~
e^{-i\omega_{\mathbf{k}_{1}}t_{1}} ~
e^{i\omega_{\mathbf{k}_{2}}t_{2}} ~
e^{-i\omega_{\mathbf{k}_{s}}(t_{2}-t_{1})} . 
\end{eqnarray} 
Here, $\hat{n}(\mathbf{x}) =  \hat{\psi}^{\dagger}(\mathbf{x}) \hat{\psi}(\mathbf{x})$ 
is the electron density operator.
For an x-ray probe pulse with  small angular spread and 
small bandwidth, the ranges of
$\mathbf{k}_{1},s_{1}$ and $\mathbf{k}_{2},s_{2}$ in which
$\rho^{X}_{\{n\},\{\bar{n}\}}$ is not negligible, are limited.
Within these ranges, the polarization vectors and the factor
$\sqrt{\omega_{\mathbf{k}_{1}}\omega_{\mathbf{k}_{2}}}$ vary
slowly with $\mathbf{k}_{1},s_{1}$ and $\mathbf{k}_{2},s_{2}$.
Here we assume that the incident x-ray pulse has a mean wave vector
$\mathbf{k}_{in}$ as the
incident photon momentum and a mean polarization vector
$\boldsymbol{\epsilon}_{\mathbf{k}_{in},s_{in}}$. 
Therefore, one can replace both
$\boldsymbol{\epsilon}_{\mathbf{k}_{1},s_{1}}$ and
$\boldsymbol{\epsilon}_{\mathbf{k}_{2},s_{2}}$ with
$\boldsymbol{\epsilon}_{\mathbf{k}_{in},s_{in}}$, and the factor
$\sqrt{\omega_{\mathbf{k}_{1}}\omega_{\mathbf{k}_{2}}}$ with
$\omega_{\mathbf{k}_{in}}$ as the photon energy of the incident central carrier frequency. 
Therefore, Eq.~(\ref{eq9}) further
simplifies to
\begin{eqnarray}\label{eq10}
\frac{dP}{d\Omega} & = & 
\frac{d\sigma_{th}}{d\Omega} ~ \int_{-\infty}^{\infty} d\gamma
\int_{-\infty}^{\infty} d\delta \int_{0}^{\infty}
d\omega_{\mathbf{k}_{s}} ~W_{\Delta
\omega}({\omega_{\mathbf{k}_{s}}}) \frac{\omega_{\mathbf{k}_{s}}
}{(2\pi \omega_{\mathbf{k}_{in}})^{2}\alpha}
~ e^{ - i\omega_{\mathbf{k}_{s}} \delta }  \nonumber \\
& & \times \sum_{I} {p_{I}}  \int d^{3}x \int d^{3}x^{\prime}
\left \langle \chi_{I},  \gamma + \frac{\delta}{2} \left| 
\left \langle \phi_{I} \left| 
~\hat{n}(\mathbf{x}^{\prime})~ \hat{U} \left(\gamma +
\frac{\delta}{2}, \gamma - \frac{\delta}{2} \right)
~\hat{n}(\mathbf{x}) ~ \right |\phi_{I} \right \rangle 
 \right| \chi_{I}, \gamma - \frac{\delta}{2} \right \rangle 
 \nonumber \\
&& \times e^{- i \mathbf{k}_{s} \cdot (
\mathbf{x}-\mathbf{x}^{\prime})}~ G^{(1)}
\left(\mathbf{x}^{\prime}, \gamma+\frac{\delta}{2};~ \mathbf{x},
\gamma-\frac{\delta}{2} \right).
\end{eqnarray}
Here, 
\begin{equation}
\frac{d\sigma_{th}}{d\Omega} = \alpha^{4} \Biggl(
\sum_{s_{s}=1}^{2}
|\boldsymbol{\epsilon}^{*}_{\mathbf{k}_{in},s_{in}} \cdot
\boldsymbol{\epsilon}_{\mathbf{k_{s}},s_{s}}|^{2} \Biggr)
\end{equation}
is the Thomson scattering cross-section of a free-electron and
$G^{(1)} \left(\mathbf{x}^{\prime}, \gamma+\frac{\delta}{2};~ \mathbf{x},
\gamma-\frac{\delta}{2} \right)$ is the first-order correlation function of the x-ray field. 
We assume that the x-ray field is a chaotic ensemble of single x-ray pulses, such that 
\begin{equation}\label{eq11}
 G^{(1)}
\left(\mathbf{x}^{\prime}, \gamma+\frac{\delta}{2};~ \mathbf{x},
\gamma-\frac{\delta}{2} \right)  = e^{- \frac{\pi \delta^{2}}{2 \delta^{2}_{c}}} 
e^{  i\omega_{\mathbf{k}_{in}} \delta }~~ 2 \pi   \alpha
I(\gamma) e^{i \mathbf{k}_{in} \cdot (
\mathbf{x}-\mathbf{x}^{\prime})},
\end{equation} 
where  
the coherence time $\delta_{c}$ is much shorter than the pulse duration
and $I(\gamma)$ is the intensity of the x-ray pulse. 
For simplicity, the object is assumed to be much smaller than the transverse coherence length.  
Note that new time variables,  $\gamma=
\frac{t_{1}+t_{2}}{2}$ and $\delta = t_{2}-t_{1}$, have been introduced in Eq.~(\ref{eq10}).

After substituting the first-order correlation function for chaotic x-ray pulses, 
the expression of the DSP can be written as
\begin{eqnarray}\label{eq12}
\frac{dP}{d\Omega} & = & 
\frac{d\sigma_{th}}{d\Omega}~
 \int_{0}^{\infty}
d\omega_{\mathbf{k}_{s}}~ W_{\Delta
\omega}({\omega_{\mathbf{k}_{s}}})~\frac{\omega_{\mathbf{k}_{s}}
}{\omega_{\mathbf{k}_{in}}}
\int_{-\infty}^{\infty} d\gamma
\frac{I(\gamma)}{\omega_{\mathbf{k}_{in}}}
\int_{-\infty}^{\infty} \frac{d\delta}{2\pi}  
C(\delta) 
~ e^{  i \omega \delta } \nonumber \\
& & \times \sum_{I} {p_{I}}  \int d^{3}x \int d^{3}x^{\prime}
\left \langle \chi_{I},  \gamma + \frac{\delta}{2} \left| 
\left \langle \phi_{I} \left| 
~\hat{n}(\mathbf{x}^{\prime})~ \hat{U} \left(\gamma +
\frac{\delta}{2}, \gamma - \frac{\delta}{2} \right)
~\hat{n}(\mathbf{x}) ~
\right |\phi_{I} \right \rangle
 \right| \chi_{I}, \gamma - \frac{\delta}{2} \right \rangle
  \nonumber \\
&& \times e^{i \mathbf{Q}\cdot(\mathbf{x} -
\mathbf{x}^{\prime})},
\end{eqnarray}
where $\omega = \omega_{\mathbf{k}_{in}}-\omega_{\mathbf{k}_{s}}$ is the photon 
energy transferred,
$\mathbf{Q} = \mathbf{k}_{in} - \mathbf{k}_{s} $ is
the photon momentum transfer,
and $C(\delta) = e^{- \frac{\pi \delta^{2}}{2 \delta^{2}_{c}}}$ defines the 
coherence function of the incident x-ray pulse.  It is important to emphasize  
that the expression given in Eq.~(\ref{eq12}) for the DSP is a key result for TRXS from an 
incoherent electronic mixture in which 
no restrictions regarding the pulse duration or coherence time of the x-ray pulse 
in comparison to the characteristic dynamical timescales of the photo-excited target system   
are employed. 
Moreover, an intuitive and straightforward  interpretation of Eq.~(\ref{eq12}) is not feasible 
as it is  intertwined with $\omega_{\mathbf{k}_{s}}$, $\gamma$, and $\delta$ variables.
In the following, we will simplify Eq.~(\ref{eq12}) considering practically relevant situations.  

\section{Results and Discussion}

\subsection{X-ray pulse short in comparison to the nuclear motion} 
Let us consider the situation where the probe x-ray pulse is sufficiently short to freeze the vibrational 
dynamics, i.e., the nuclear motion is assumed to be
much slower than the x-ray pulse duration. Within this assumption, 
after collecting all the 
$\gamma$-dependent phases of the vibrational wave packets 
together with the $I(\gamma)$,  the $\gamma$-dependent integral in Eq.~(\ref{eq12}) can
be performed  such that 
 \begin{eqnarray}\label{eq13}
\frac{dP}{d\Omega} & = &  \mathcal{F} 
\frac{d\sigma_{th}}{d\Omega}~
 \int_{0}^{\infty}
d\omega_{\mathbf{k}_{s}}~ W_{\Delta
\omega}({\omega_{\mathbf{k}_{s}}})~\frac{\omega_{\mathbf{k}_{s}}
}{\omega_{\mathbf{k}_{in}}}
\int_{-\infty}^{\infty} \frac{d\delta}{2\pi}  
C(\delta) 
~ e^{  i \omega \delta }  \sum_{I} \sum_{J; \mu}  {p_{I}}   
\int d^{3}x \int d^{3}x^{\prime}  \nonumber \\
& & \times 
\bigg \langle \chi_{I},  \tau_{d} + \frac{\delta}{2}  \bigg | 
 \bigg \langle \phi_{I} \bigg | \hat{n}(\mathbf{x}^{\prime})
 \bigg | \phi_{J} \bigg \rangle \bigg | \chi_{J; \mu} \bigg \rangle 
e^{-i E_{J; \mu} \delta}  
\bigg \langle \chi_{J; \mu} \bigg | \bigg \langle \phi_{J} \bigg |  
\hat{n}(\mathbf{x})  \bigg | \phi_{I}  \bigg \rangle 
\bigg | \chi_{I}, \tau_{d} - \frac{\delta}{2} \bigg \rangle \nonumber \\
&& \times e^{i \mathbf{Q}\cdot(\mathbf{x} -
\mathbf{x}^{\prime})}. 
\end{eqnarray}
Here, $\mathcal{F}$ is the fluence of the probe pulse 
(in units of number of photons per area) and
$\tau_{d}$ is the pump-probe delay time. 
Note that if one had allowed for any electronic coherences in the superposition of electronic states, then the $\gamma$-dependent integration would have eliminated any contributions from pairs of different electronic states, because an x-ray probe pulse duration of femtoseconds is long in comparison to the typical sub-fs electronic timescales of interest here.

After substituting the expression for the vibrational wave packet, Eq.~(\ref{eq14}), 
the expression for the DSP becomes 
\begin{eqnarray}\label{eq15a}
\frac{dP}{d\Omega} & = &  \mathcal{F} 
\frac{d\sigma_{th}}{d\Omega}~
 \int_{0}^{\infty}
d\omega_{\mathbf{k}_{s}}~ W_{\Delta
\omega}({\omega_{\mathbf{k}_{s}}})~\frac{\omega_{\mathbf{k}_{s}}
}{\omega_{\mathbf{k}_{in}}}
\int_{-\infty}^{\infty} \frac{d\delta}{2\pi}  
C(\delta) 
~ e^{  i \omega \delta }   
\nonumber \\
& & \times \sum_{I; \nu, \xi} \sum_{J; \mu}   {p_{I}}  
C_{I; \nu}^{*}  C_{I; \xi} \int d^{3}x \int d^{3}x^{\prime}
\langle \chi_{I; \nu}|  
~ \langle \phi_{I} | \hat{n}(\mathbf{x}^{\prime})
 |\phi_{J} \rangle~
 | \chi_{J; \mu} \rangle  
\langle \chi_{J; \mu} | ~\langle \phi_{J}|  
\hat{n}(\mathbf{x})  |\phi_{I}  \rangle  ~| \chi_{I; \xi} \rangle
 \nonumber \\
&& \times ~ e^{i (E_{I; \nu}-E_{I; \xi}) \tau_{d}} 
e^{-i (2E_{J; \mu}-E_{I; \nu}-E_{I; \xi}) \frac{\delta}{2}} e^{i \mathbf{Q}\cdot(\mathbf{x} -
\mathbf{x}^{\prime})}. 
\end{eqnarray}

Now the $\delta$-integral can be performed straightaway, 
yielding  another Gaussian as a function of the coherence time:
\begin{equation}\label{eq15}
\int_{-\infty}^{\infty} \frac{d\delta}{2\pi} e^{- \frac{\pi \delta^{2}}{2 \delta^{2}_{c}}}
e^{  -i (\omega_{\mathbf{k}_{s}}-\omega_{\mathbf{k}_{in}}) \delta } 
e^{-i (2E_{J; \mu}-E_{I; \nu}-E_{I; \xi}) \frac{\delta}{2}} 
= \frac{\delta_{c}}{\pi \sqrt{2}} 
e^{- \frac{\delta^{2}_{c}}{2 \pi}(\omega_{\mathbf{k}_{in}}- \omega_{\mathbf{k}_{s}} + \tilde{E}_{I} 
- E_{J; \mu})^{2}},
\end{equation} 
where $(E_{I; \nu}+E_{I; \xi})/2$ has been replaced with  the mean energy 
$\tilde{E}_{I}$ of the vibrational wave packet in the $I$-th electronic state. 
As the x-ray pulse duration is short enough to freeze the nuclear motion, 
$|\tilde{E}_{I} - E_{I; \mu}| \ll 1/\delta_{c}$ holds and  
$\textrm{exp}[i (E_{I;\nu} - \tilde{E}_{I}) \delta_{c}/2]$ may be approximated by unity.

The typical situation is that the TRXS experiment does not use 
an energy-resolving scattering 
detector, i.e.,  
$W_{\Delta \omega}({\omega_{\mathbf{k}_{s}}}) = 1$. 
Then  the energy integral can also be performed:  
\begin{equation}\label{eq16}
\int_{0}^{\infty} d\omega_{\mathbf{k}_{s}} \omega_{\mathbf{k}_{s}}  \frac{\delta_{c}}{\pi \sqrt{2}} 
e^{- \frac{\delta^{2}_{c}}{2 \pi}(\omega_{\mathbf{k}_{in}}- \omega_{\mathbf{k}_{s}} + \tilde{E}_{I} - E_{J; \mu})^{2}} \simeq \omega_{\mathbf{k}_{in}}.
\end{equation} 
To obtain this result, 
two assumptions are made: (i) 
only electronic and vibrational states contribute  to the total scattering signal 
such that $|\tilde{E}_{I} - E_{J; \mu}| \ll \omega_{\mathbf{k}_{in}}$, i.e., the scattering process does not lead to an energy transfer anywhere close to the incoming photon energy~\cite{santra2008concepts, slowik2014incoherent}, and (ii) 
the width of the Gaussian in Eq.~(\ref{eq16}) must be small in comparison to $ \omega_{\mathbf{k}_{in}}$, i.e., the incoming x-ray pulse must be sufficiently monochromatic such that 
$\delta_{c} ~ \omega_{\mathbf{k}_{in}} \gg 1 $. 
Note that the  Waller-Hartree approximation is used in performing the  $\omega_{\mathbf{k}_{s}}$-integral~\cite{slowik2014incoherent, waller1929intensity}. 

After combining the results obtained after performing the $\delta$- and $\omega_{\mathbf{k}_{s}}$-integrals in 
Eqs.~(\ref{eq15})  and~(\ref{eq16}), respectively,
the expression for the DSP reduces to
\begin{eqnarray}\label{eq017}
\frac{dP}{d\Omega}  & =  & 
\frac{dP_{e}}{d\Omega}~
\sum_{I}  \sum_{J; \mu}  {p_{I}}   
 \int d^{3}x \int d^{3}x^{\prime} 
\nonumber \\
& & \times 
\langle \chi_{I}, \tau_{d}|  
~ \langle \phi_{I} | \hat{n}(\mathbf{x}^{\prime})
 |\phi_{J} \rangle~
 | \chi_{J; \mu} \rangle  
\langle \chi_{J; \mu} | ~\langle \phi_{J}|  
\hat{n}(\mathbf{x})  |\phi_{I}  \rangle  ~| \chi_{I}, \tau_{d} \rangle
e^{i \mathbf{Q}\cdot(\mathbf{x} -
\mathbf{x}^{\prime})},
\end{eqnarray} 
where $\frac{dP_{e}}{d\Omega} =  \mathcal{F} 
\frac{d\sigma_{th}}{d\Omega}$ is the DSP from a free electron.  
Note that first quantization for the nuclear degrees of freedom is employed
here, i.e.,  $\langle \chi_{I}| \langle \phi_{I} | \hat{n}|\phi_{J} \rangle | \chi_{J} \rangle$  
implies an integration over $\mathbf{R}$, i.e., 
$\int d\mathbf{R} ~\chi_{I}^{*}(\mathbf{R})~ \langle \phi_{I} | \hat{n}|\phi_{J} \rangle(\mathbf{R}) 
~\chi_{J}(\mathbf{R})$, where $\langle \phi_{I} | \hat{n}|\phi_{J} \rangle(\mathbf{R})$ acts as an operator in $\mathbf{R}$-space.
  
Now let us consider TRXS 
from a molecule consisting of a heavy element like iodine. 
This corresponds to the situation of the recent  TRXS experiment~\cite{glownia2016self} 
on the iodine molecule. 
As there are many electrons in iodine, one gets 
a large cross section if the electrons don't change their state in the x-ray scattering process.
So it is appropriate to assume that the dominating contribution comes 
from electronic terms where $I = J$, i.e.,  electronically elastic scattering. 
Thus, under this assumption the DSP is given by  
\begin{equation}\label{eq17}
\frac{dP}{d\Omega}   =   \frac{dP_{e}}{d\Omega}~
\sum_{I}   p_{I} 
\langle \chi_{I}, \tau_{d}| |f_{I}(\mathbf{Q})|^{2} 
| \chi_{I}, \tau_{d} \rangle.
\end{equation} 
Here, $ f_{I}(\mathbf{Q})  = \int d^{3}x \langle \phi_{I}|  
\hat{n}(\mathbf{x})  |\phi_{I}  \rangle e^{i \mathbf{Q} \cdot \mathbf{x}} =   
\int d^{3}x \rho_{I}(\mathbf{x}) e^{i \mathbf{Q} \cdot \mathbf{x}} $  is  the electronic 
form factor.
The completeness relation in $\mathbf{R}$-space,
\begin{equation}\label{eq18}
\sum_{\mu} | \chi_{J; \mu} \rangle \langle \chi_{J; \mu}|  = \mathds{1}_{\mathbf{R}},
\end{equation}  
was used to obtain Eq.~(\ref{eq17}) from Eq.~(\ref{eq017}). 

Now if the vibrational distribution 
$|\chi_{I}(\mathbf{R}, \tau_{d})|^{2}$ 
is sufficiently narrow, i.e., the electronic form factor squared must be approximately 
constant in the vicinity of the first moment of the vibrational distribution where
 the extension of that vicinity is given by the width of the vibrational distribution,
 then the main quantity of Eq.~(\ref{eq17}) may be simplified as
$\langle \chi_{I}, \tau_{d}| |f_{I}(\mathbf{Q})|^{2} 
| \chi_{I}, \tau_{d} \rangle \approx \langle \chi_{I}, \tau_{d}| \chi_{I}, \tau_{d} \rangle  
|f_{I}(\mathbf{Q}, \mathbf{R}^{(I)}_{\tau_{d}})|^{2} =  |f_{I}(\mathbf{Q}, \mathbf{R}_{\tau_{d}}^{(I)})|^{2}$ with $\mathbf{R}_{\tau_{d}}^{(I)} = 
\langle \chi_{I}, \tau_{d}| \mathbf{R} |\chi_{I}, \tau_{d} \rangle$.  
Using this approximation, Eq.~(\ref{eq17}) may be written as
\begin{equation}\label{eq17a}
\frac{dP}{d\Omega}   =   \frac{dP_{e}}{d\Omega}~
\sum_{I}   p_{I} 
|f_{I}(\mathbf{Q}, \mathbf{R}_{\tau_{d}}^{(I)})|^{2}. 
\end{equation}

Let us compare this result to the expression employed in Ref.~\cite{glownia2016self}. 
To this end, we restrict the summation over $I$ in Eq.~(\ref{eq17a}) to only two 
electronic states: the ground electronic state ($g$ in the notation of Ref.~\cite{glownia2016self}) 
and the first excited electronic state ($e$). Thus, 
\begin{equation}\label{eq17b}
\frac{dP}{d\Omega}   =  \frac{dP_{e}}{d\Omega}~
\left[ a |f^{(e)}(\mathbf{Q}, \mathbf{R}^{(e)}_{\tau_{d}})|^{2} 
+ (1-a) |f^{(g)}(\mathbf{Q}, \mathbf{R}^{(g)}_{\tau_{d}})|^{2} \right],
\end{equation}
where $a$ represents the population in the excited electronic state. Expressed in words,
the scattering signal in the situation considered is obtained by incoherently averaging
the differential scattering probabilities associated with the two electronic states $g$ and $e$.
By contrast, Eq. (4) in Ref.~\cite{glownia2016self}) suggests that 
the x-ray scattering intensity at the detector is proportional to
$|a f^{(e)} + (1-a) f^{(g)}|^{2}$, which represents a coherent average of the scattering
amplitudes and, as demonstrated here, is not applicable to an incoherent electronic mixture. 
Interestingly, as shown in Ref.~\cite{dixit2012}, such a coherent average is generally not 
even applicable when the electronic superposition is perfectly coherent.
[Note that the electronic form factors $f_{I}$ in Eq.~(\ref{eq17a}) refer to the entire 
system considered. In the case of a gas-phase system, such as that considered in 
Ref.~\cite{glownia2016self}, one must average Eq.~(\ref{eq17a}) over all molecular positions.
As a consequence, molecule--molecule interference terms drop out. For that reason,
the electronic form factors in Eq.~(\ref{eq17b}) refer to individual molecules. 
More on this in Sec.~\ref{crystal}.]

Our present finding does not support the idea of heterodyne detection of the scattering signal
obtained from an incoherent electronic mixture in gas-phase photo-excited molecules, which was the 
key idea to analyze the recent experimental work in Ref.~\cite{glownia2016self};  
our present result is thus consistent with the brief remark made by 
Mukamel and co-workers~\cite{bennett2017comment} (see also Refs.~\cite{cao1998ultrafast, bratos2002}).
As reflected in Eq.~(\ref{eq17}), the total scattering signal from an incoherent electronic mixture 
consists of an incoherent sum of individual scattering patterns
obtained from each electronic state and weighted by an appropriate nuclear wave-packet
density. The signal obtained from each electronic state is sensitive to
vibrational coherence in each electronic state as a function of the pump-probe delay time.

The  fractional contribution to  the scattering signal 
from an  excited electronic state is precisely 
of the order of the associated  excitation probability, as one would expect. 
Ignoring higher-order terms of the excitation probability
is not required in the present case, which is in contrast to the 
analysis used in Ref.~\cite{glownia2016self}, where the square of the  excitation probability
was ignored to explain an enhancement of the excited electronic state scattering signal using the 
heterodyne detection concept.

It is very important to stress that assigning an additional degree of freedom corresponding to 
 the pump-probe delay time to the total static electron density when transiting 
from static x-ray scattering to TRXS and writing the total time-dependent 
electron density as a sum of electron densities, one for  each electronic state,  
is not a correct approach and leads to a wrong interpretation of TRXS.  
But this brings us to another important question: If heterodyne detection is not feasible in the case 
of an incoherent electronic mixture in gas-phase photo-excited molecules,
then what justifies the extensively applied heterodyne detection for analyzing  
the total scattering signal from an electronically excited 
crystal~\cite{elsaesser2014perspective, zamponi2012ultrafast}? 
We will answer this question in the next subsection. 

\subsection{Time-resolved x-ray scattering from an electronically excited crystal}
\label{crystal}
In order to develop  a consistent quantum theory-based formalism for 
ultrafast x-ray scattering from an electronically excited crystal we employ the following  assumptions: 
\begin{enumerate}
\item There is no electronic coherence. 
\item The unit cells may be assumed to be independent of one another (independent-unit-cell model). 
\item Each unit cell is either in its electronic ground state or in a well-defined excited state. In other words, only two eigenstates are considered. 
\item In the ground electronic state, the vibrational probability distribution is assumed to be stationary. 
In the excited state, the vibrational distribution is nonstationary.  
\end{enumerate}

In order to obtain the TRXS signal from an electronically excited crystal,  
the key  quantity from Eq.~(\ref{eq17}) is 
\begin{equation}\label{eq26}
S =   \sum_{I} p_{I} 
\langle \chi_{I}, \tau_{d}|  
 ~|f_{I}(\mathbf{Q})|^{2}  ~| \chi_{I},  \tau_{d} \rangle 
 = \sum_{I} p_{I} \int d\mathbf{R}~ |\chi_{I} (\mathbf{R}, \tau_{d})|^{2} 
~\left| \int d^{3}x~\rho_{I}(\mathbf{x}, \mathbf{R})~e^{i \mathbf{Q} \cdot \mathbf{x}} \right|^{2}.  
\end{equation} 
We now use the index $i$ to refer to the $i$-th unit cell. 
Following the procedure described in 
Refs.~\cite{son2011multiwavelength, son2013determination}, the global electronic configuration index
$I$ is given by $I = (I_{1}, I_{2}, \ldots, I_{i}, \ldots),$ where each $I_{i}$ takes on only two values 
(0 and 1, or ground state (GS) and excited state (ES)). 
By assumption of the independent-unit-cell model, 
\begin{equation}\label{eq27}
\underbrace{p_{I} |\chi_{I} (\mathbf{R}, \tau_{d})|^{2}}_{\textrm{entire crystal}} 
= \prod_{i}  \underbrace{p_{I_{i}} |\chi_{I_{i}} (\mathbf{R}_{i}, \tau_{d})|^{2}}_{\textrm{$i$-th unit cell}}.  
\end{equation} 
Here, $\mathbf{R}_{i}$ represents the nuclear positions in the $i$-th unit cell;
$p_{I_{i}}$ is the associated electronic population and $\chi_{I_{i}} (\mathbf{R}_{i}, \tau_{d})$ 
is the associated vibrational wave function.  
Now let us express the total electron
density in terms of the electron density of individual  unit cells:  
\begin{equation}\label{eq28}
 \int d^{3}x~\rho_{I}(\mathbf{x}, \mathbf{R})~e^{i \mathbf{Q} \cdot \mathbf{x}}
 = \sum_{i}   \int d^{3}x  \rho_{I_{i}}(\mathbf{x}, \mathbf{R}_{i})~e^{i \mathbf{Q} \cdot \mathbf{x}},
\end{equation} 
where $\rho_{I_{i}}(\mathbf{x}, \mathbf{R}_{i})$ is the electron density in the $i$-th unit cell,
which  depends on the electronic state of the $i$-th unit cell and on the nuclear positions in that unit cell. 
Let $\mathbf{r}_{i}$ be the real-space lattice vector for the position of the $i$-th 
unit cell. Using this, we write the electron position for the $i$-th unit cell as 
$\mathbf{x}_{i} = \mathbf{x} - \mathbf{r}_{i}$. Hence,  the above equation becomes 
\begin{equation}\label{eq29}
 \int d^{3}x~\rho_{I}(\mathbf{x}, \mathbf{R})~e^{i \mathbf{Q} \cdot \mathbf{x}}
 = \sum_{i}  \left[  \int d^{3}x_{i}  \rho_{I_{i}}(\mathbf{x}_{i}+\mathbf{r}_{i}, \mathbf{R}_{i})~e^{i  \mathbf{Q} \cdot \mathbf{x}_{i} } \right] e^{i \mathbf{Q}  \cdot \mathbf{r}_{i} }.
\end{equation} 
Let 
$F_{i}(\mathbf{Q}) = \int d\mathbf{x}_{i} ~\rho_{I_{i}}(\mathbf{x}_{i}+\mathbf{r}_{i}, \mathbf{R}_{i})~e^{i  \mathbf{Q} \cdot \mathbf{x}_{i} }$ 
be the structure factor of the $i$-th unit cell in the electronic state $I_{i}$ for a given $\mathbf{R}_{i}$. 
The dependence of $F_{i}(\mathbf{Q})$ on $I_{i}$ and $\mathbf{R}_{i}$ must not be forgotten. 
So, Eq.~(\ref{eq29}) reduces to 
\begin{equation}\label{eq31}
 \int d^{3}x~\rho_{I}(\mathbf{x}, \mathbf{R})~e^{i \mathbf{Q} \cdot \mathbf{x}} = 
\sum_{i} F_{i}(\mathbf{Q}) e^{i  \mathbf{Q} \cdot \mathbf{r}_{i} }.
\end{equation} 

Thus, after collecting results from Eqs.~(\ref{eq27})--(\ref{eq31}), Eq.~(\ref{eq26}) is written as 
\begin{equation}\label{eq32}
S = \sum_{I_{1}, I_{2}, \ldots} \prod_{i} p_{I_{i}} 
\int d\mathbf{R}_{i} ~|\chi_{I_{i}} (\mathbf{R}_{i}, \tau_{d})|^{2}
\sum_{i^{\prime}, j^{\prime}} F_{i^{\prime}}(\mathbf{Q}) F_{j^{\prime}}^{*}(\mathbf{Q}) 
~e^{i  \mathbf{Q} \cdot ( \mathbf{r}_{i^{\prime}} - \mathbf{r}_{j^{\prime}} )}.
\end{equation} 
We must distinguish between terms where $i^{\prime} =  j^{\prime} $ and terms where 
$i^{\prime} \neq  j^{\prime}$, i.e., 
\begin{equation}\label{eq33}
S =  \sum_{i^{\prime}} S_{i^{\prime} i^{\prime}} + 
\sum_{\substack{i^{\prime}, ~ j^{\prime} \\
                               i^{\prime} \neq  j^{\prime} }}
S_{i^{\prime} j^{\prime}}~e^{i  \mathbf{Q} \cdot ( \mathbf{r}_{i^{\prime}} - \mathbf{r}_{j^{\prime}} )},
\end{equation} 
where
\begin{eqnarray}\label{eq34}
S_{i^{\prime} i^{\prime}}  & = & \sum_{I_{1}, I_{2}, \ldots} \prod_{i} p_{I_{i}}  
\int d\mathbf{R}_{i} ~|\chi_{I_{i}} (\mathbf{R}_{i}, \tau_{d})|^{2}
| F_{i^{\prime}}(\mathbf{Q})|^{2} \nonumber \\
& = & \prod_{i \neq i^{\prime}} \sum_{I_{i}} p_{I_{i}}    \int d\mathbf{R}_{i} ~|\chi_{I_{i}} 
(\mathbf{R}_{i}, \tau_{d})|^{2}
\left( \sum_{I_{i^{\prime}}} p_{I_{i^{\prime}}}  
\int d\mathbf{R}_{i^{\prime}} ~|\chi_{I_{i^{\prime}}} (\mathbf{R}_{i^{\prime}}, \tau_{d})|^{2}
| F_{i^{\prime}}(\mathbf{Q})|^{2} \right).
\end{eqnarray} 
We assume that $\langle \chi_{I}|\chi_{I} \rangle_{\mathbf{R}} = 1,$
i.e., the vibrational wave function for all nuclei is normalized. 
Since we further assume that each $\chi_{I}$ factorizes into factors $\chi_{I_{i}}(\mathbf{R}_{i}, \tau_{d}),$ one for each unit cell, this is consistent with assuming that each 
 $\chi_{I_{i}}(\mathbf{R}_{i}, \tau_{d})$ is normalized. This means
 \begin{equation}\label{eq35}
\prod_{i \neq i^{\prime}} \sum_{I_{i}} p_{I_{i}}   
\underbrace{ \int d\mathbf{R}_{i} ~|\chi_{I_{i}} (\mathbf{R}_{i}, \tau_{d})|^{2}}_{= 1}
= \prod_{i \neq i^{\prime}} \sum_{I_{i}} p_{I_{i}}  = 1.
\end{equation} 
Here we exploited that irrespective of whether one allows only two or more electronic states, 
$ \sum_{I_{i}} p_{I_{i}}  = 1$. Hence, 
\begin{eqnarray}\label{eq36}
S_{i^{\prime} i^{\prime}}  & = &
 \sum_{I_{i^{\prime}}} p_{I_{i^{\prime}}}   
\int d\mathbf{R}_{i^{\prime}} ~|\chi_{I_{i^{\prime}}} (\mathbf{R}_{i^{\prime}}, \tau_{d})|^{2}
| F_{i^{\prime}}(\mathbf{Q})|^{2} \nonumber \\
& = & (1-\eta) \int d\tilde{\mathbf{R}} ~|\chi_{GS}(\tilde{\mathbf{R}})|^{2} 
~|F^{GS}(\mathbf{Q}, \tilde{\mathbf{R}})|^{2}
+ \eta \int d\tilde{\mathbf{R}} ~|\chi_{ES}(\tilde{\mathbf{R}}, \tau_{d})|^{2} 
~|F^{ES}(\mathbf{Q}, \tilde{\mathbf{R}})|^{2}. 
\end{eqnarray} 
In this expression, $\tilde{\mathbf{R}}$ denotes the nuclear coordinates in the unit cell. Because the scattering contributions $S_{i^{\prime} i^{\prime}}$ depend on the 
electronic state of unit cell $i^{\prime}$ (and on the vibrational state associated with the electronic state), but not on the actual position of the unit cell, we may employ a generic 
$\tilde{\mathbf{R}}$ rather than ${\mathbf{R}_{i^{\prime}}}$. 
Here, the connection of $p_{I_{i}}$ in the above equation with the notation used by 
Elsaesser and co-workers~\cite{elsaesser2014perspective, zamponi2012ultrafast}  
has been made via  
\begin{equation}\label{eq37}
 p_{I_{i}}  = 
 \begin{cases}
\eta, ~~\textrm{if}~~I_{i} = 1~~(\textrm{ES})\\
(1-\eta), ~~\textrm{if}~~I_{i} = 0~~(\textrm{GS})\\
\end{cases}   
\end{equation} 

Now, let us consider the other part of Eq.~(\ref{eq33}): 
\begin{eqnarray}\label{eq38}
S_{i^{\prime} j^{\prime}}  & = & \sum_{I_{1}, I_{2}, \ldots} \prod_{i} p_{I_{i}}  
\int d\mathbf{R}_{i} ~|\chi_{I_{i}} (\mathbf{R}_{i}, \tau_{d})|^{2}
F_{i^{\prime}}(\mathbf{Q})  F_{j^{\prime}}^{*}(\mathbf{Q})  \nonumber \\
& = & \left(  \sum_{I_{i^{\prime}}} p_{I_{i^{\prime}}}    
\int d\mathbf{R}_{i^{\prime}} ~|\chi_{I_{i^{\prime}}} (\mathbf{R}_{i^{\prime}}, \tau_{d})|^{2}
F_{i^{\prime}}(\mathbf{Q}) \right) 
 \left(  \sum_{I_{j^{\prime}}} p_{I_{j^{\prime}}}    
\int d\mathbf{R}_{j^{\prime}} ~|\chi_{I_{j^{\prime}}} (\mathbf{R}_{j^{\prime}}, \tau_{d})|^{2}
F_{j^{\prime}}^{*}(\mathbf{Q}) \right) \nonumber \\
& = & \left[  
(1-\eta) \int d\tilde{\mathbf{R}} ~|\chi_{GS}(\tilde{\mathbf{R}})|^{2} 
~F^{GS}(\mathbf{Q}, \tilde{\mathbf{R}})
+ \eta \int d\tilde{\mathbf{R}} ~|\chi_{ES}(\tilde{\mathbf{R}}, \tau_{d})|^{2} 
~F^{ES}(\mathbf{Q}, \tilde{\mathbf{R}})
\right] \nonumber \\
&& \times 
 \left[  
(1-\eta) \int d\tilde{\mathbf{R}} ~|\chi_{GS}(\tilde{\mathbf{R}})|^{2} 
~(F^{GS}(\mathbf{Q}, \tilde{\mathbf{R}}))^{*}
+ \eta \int d\tilde{\mathbf{R}} ~|\chi_{ES}(\tilde{\mathbf{R}}, \tau_{d})|^{2} 
~(F^{ES}(\mathbf{Q}, \tilde{\mathbf{R}}))^{*}
\right]. \nonumber \\
\end{eqnarray} 
This motivates introducing effective structure factors that are averaged over nuclear coordinates:
$F^{GS}_{eff}(\mathbf{Q}) = \int d\tilde{\mathbf{R}} ~|\chi_{GS}(\tilde{\mathbf{R}})|^{2} 
~F^{GS}(\mathbf{Q}, \tilde{\mathbf{R}}),$
and
$F^{ES}_{eff}(\mathbf{Q}, \tau_{d}) = 
\int d\tilde{\mathbf{R}} ~|\chi_{ES}(\tilde{\mathbf{R}}, \tau_{d})|^{2} 
~F^{ES}(\mathbf{Q}, \tilde{\mathbf{R}})$.
Using  these  two effective structure factors, we arrive at
\begin{equation}\label{eq40}
S_{i^{\prime} j^{\prime}}  = 
| (1-\eta) F^{GS}_{eff}(\mathbf{Q}) +  \eta F^{ES}_{eff}(\mathbf{Q}, \tau_{d})|^{2}.
\end{equation} 

Using the results from Eqs.~(\ref{eq33}), (\ref{eq36}), and ~(\ref{eq40}), it follows that  
\begin{eqnarray}\label{eq41}
S   & = & 
\sum_{i} \left[ (1-\eta) \langle (F^{GS})^{2} \rangle  + \eta \langle (F^{ES})^{2} \rangle  \right]
+ \sum_{i \neq j} \left| (1-\eta) F^{GS}_{eff}(\mathbf{Q}) + \eta F^{ES}_{eff}(\mathbf{Q}, \tau_{d}) \right|^{2} 
e^{i  \mathbf{Q} \cdot (\mathbf{r}_{i} - \mathbf{r}_{j})} \nonumber \\
& = &  \left| (1-\eta) F^{GS}_{eff}(\mathbf{Q}) + \eta F^{ES}_{eff}(\mathbf{Q}, \tau_{d}) \right|^{2}  
\left| \sum_{i} e^{i  \mathbf{Q} \cdot \mathbf{r}_{i}} \right|^{2} \nonumber \\
&& + \sum_{i} \left[ (1-\eta) \langle (F^{GS})^{2} \rangle  + \eta \langle (F^{ES})^{2} \rangle 
- |(1-\eta) F^{GS}_{eff}(\mathbf{Q}) + \eta F^{ES}_{eff}(\mathbf{Q}, \tau_{d})|^{2} \right], 
\end{eqnarray} 
where the short-hand notations  
$\langle (F^{GS})^{2} \rangle  = 
\int d\tilde{\mathbf{R}} ~|\chi_{GS}(\tilde{\mathbf{R}})|^{2} 
~|F^{GS}(\mathbf{Q}, \tilde{\mathbf{R}})|^{2}$ 
and 
$\langle (F^{ES})^{2} \rangle  = 
\int d\tilde{\mathbf{R}} ~|\chi_{ES}(\tilde{\mathbf{R}}), \tau_{d}|^{2} 
~|F^{ES}(\mathbf{Q}, \tilde{\mathbf{R}})|^{2}$ are used. 
The second term on the right-hand side of the second equality sign in  Eq.~(\ref{eq41}) gives rise to diffuse scattering, whereas the first term contains the lattice sum $\sum_{i}e^{i  \mathbf{Q} \cdot \mathbf{r}_{i}},$ which gives rise to Bragg scattering at reciprocal lattice vectors $\mathbf{Q} = \mathbf{G}_{hkl}$. Hence, the quantum theory developed here gives Bragg peak strengths proportional to 
$|(1-\eta) F^{GS}_{eff}(\mathbf{G}_{hkl}) + \eta F^{ES}_{eff}(\mathbf{G}_{hkl}, \tau_{d})|^{2}$, which 
is consistent with the expression used by Elsaesser and co-workers 
for the heterodyne detection in TRXS from an electronically excited crystal~\cite{elsaesser2014perspective, zamponi2012ultrafast}. 
Finally, note that Eq.~(\ref{eq33}) applies also to the situation of TRXS from a photo-excited gas-phase sample. In that case, $\mathbf{r}_{i^{\prime}}$ and $\mathbf{r}_{j^{\prime}}$ refer to the positions of the individual gas-phase molecules. After averaging over those positions, all contributions from $i \ne j$ disappear. As a consequence, the scattering signal is simply proportional to Eq.~(\ref{eq36}): The expression required to describe the recent TRXS experiment on photo-excited molecular iodine~\cite{glownia2016self} is an incoherent sum of two scattering patterns, one corresponding to the ground electronic state and the other to the excited electronic state.

\section{Conclusion}
The present work is focused on a rigorous, formal understanding of TRXS 
from an incoherent  electronic mixture in gas-phase 
photo-excited molecules and electronically excited crystals.  
The considered electronic mixture 
has no coherence between electronic states, but has perfect vibrational coherence. 
In the case of gas-phase 
photo-excited molecules, the total scattering signal consists of an incoherent 
sum of the signal associated with each electronic state weighted by the corresponding 
nuclear wave-packet densities. We find that there is no possibility of 
heterodyning of the signal related to different electronic states.  
Our finding remains unchanged even if we consider that the 
coherence time and pulse duration of the x-ray pulse are long in comparison to 
the timescale of the electronic motion but short in comparison to 
timescale of  the vibrational motion.  
This conclusion is in contrast with recent experimental work where  
heterodyne detection was used to analyze the total scattering signal~\cite{glownia2016self}. 
Also, a gas-phase sample of aligned molecules does not give rise to any molecule-molecule interferences, as the center-of-mass position vectors of the molecules remain completely random relative to each other. Therefore, alignment doesn't change this conclusion and rules out 
the feasibility of heterodyne detection.  
In the case of an electronically excited crystal, the total signal contains  
interference between signals arising from different electronic states,
as a consequence of interfering contributions from different unit cells,
even when vibrations are systematically taken into consideration, and shows 
the feasibility of the heterodyne detection. 
Note that we can derive the expression used by Elsaesser and co-workers 
under the assumption of the independent-unit-cell approximation; 
that approximation cannot in general be expected to be accurate for a real solid.
The conceptual feasibility of heterodyne detection in a photo-excited crystal is 
 due to the periodic nature of the crystal. 
 In a gas-phase sample, the interferences required for heterodyning disappear. 
 We believe that our present findings for TRXS from photo-excited gas-phase  molecules and crystals  
 will help to develop a better understanding of current and future TRXS experiments.

\section{Acknowledgements}
We thank David Reis for helpful comments on our manuscript. 
G.D. acknowledges support from a Ramanujan fellowship (SB/S2/ RJN-152/2015).

\end{document}